# Stability of the proton-to-electron mass ratio

A. Shelkovnikov[†], R.J. Butcher[‡], C. Chardonnet and A. Amy-Klein*,

Laboratoire de Physique des Lasers, UMR CNRS 7538,

Institut Galilée, Université Paris 13, 99, ave J.-B. Clément – 93430 Villetaneuse – France

**Abstract**

We report a limit on the fractional temporal variation of the proton-to-electron mass ratio as $\frac{1}{(m_P/m_e)}\frac{\partial}{\partial t}(m_P/m_e) = (-3.8 \pm 5.6) \times 10^{-14}\ yr^{-1}$, obtained by comparing the frequency of a rovibrational transition in $SF_6$ with the fundamental hyperfine transition in Cs. The $SF_6$ transition was accessed using a $CO_2$ laser to interrogate spatial 2-photon Ramsey fringes. The atomic transition was accessed using a primary standard controlled with a Cs fountain. This result is direct and model-free.

Pacs numbers: 06.30.Ft, 06.20.Jr, 42.62.Eh, 33.20.Ea

Data obtained from high precision frequency metrology and from observational astronomy are now of such quality that they open seriously the questions of the stability of the fundamental constants. Experiments are all the more important because there is no established theory uniting the fundamental forces. Theoretical models agree that the Equivalence Principle of General Relativity is abandoned at some level and interactions thus become a function of both time and space. The two variables may be exploited in current experiments, either using the high precision of frequency metrology [1-4] or astronomical distances and, therefore, times [5]. From a growing number of theoretical papers, three reviews might be mentioned [6-8]. In the large majority of laboratory experiments two atomic clocks are compared. The fine structure constant, α, and the Rydberg, Ry, are among the significant parameters [1-4]. Data are also available from the geological record [6-9] and from astronomy [5,10,11]. We present here the first experimental comparison of a molecular clock to an atomic clock, which gives a direct line to the proton-to-electron mass ratio.

The principle of the experiment is simple. We measure the frequency of a molecular transition in $SF_6$, interrogated by a carbon dioxide laser, relative to an atomic transition in Cs. Because these are respectively vibration-rotation and hyperfine transitions, the frequencies scale as: $\nu(SF_6) = K_1 \left(\frac{m_e}{m_p}\right)^{1/2} Ry$, and $\nu(Cs) = K_2 \left(\frac{\mu_{Cs}}{\mu_B}\right) \alpha^2 F(\alpha) Ry$, giving the dependence :

$$\frac{\nu(SF_6)}{\nu(Cs)} = K \left(\frac{m_e}{m_p}\right)^{1/2} \left(\frac{\mu_B}{\mu_{Cs}}\right) \alpha^{-2} (F(\alpha))^{-1},$$ [7, 12]. The Ks are constants, $\mu_{Cs}$ is the magnetic dipole of the Cs nucleus, $\mu_B$ is the Bohr magneton, $F(\alpha)$ a dimensionless function accounting for relativistic effects in Cs, where its dependence on α is the power of 0.83. One of the obvious limits on the stability of the $SF_6$ frequency measurement is thus the stability of the ratio $m_p$:$m_e$.

The experimental arrangement is shown in Figure 1 and is composed of two parts: the $SF_6$ high-resolution spectrometer and the measurement chain [13, 14]. Briefly, a carbon dioxide laser at 10 μm (28.4 THz) is used to record two-photon Ramsey fringes on a supersonic beam of $SF_6$. The transition is P(4) $E^0$ in the $2\nu_3$ band. A folded cavity is used to provide the two phase-coherent stationary waves of the Ramsey spatial interferometer. This ensures that the central fringe is in exact coincidence with the two-photon resonance. The Ramsey fringe signal is detected by stimulated emission from the upper energy level to the intermediate rovibrational level in a separate Fabry-Perot cavity. The beam velocity is 400 m/s, and the distance between the two interaction zones is 1 m, leading to a fringe periodicity of 200 Hz [13]. The carbon dioxide laser is offset phase-locked to a second carbon dioxide laser, which itself is stabilised on a 2-photon transition in $SF_6$, FWHM 40 kHz. The $SF_6$, is contained in a Fabry-Perot cavity, FWHM 1 MHz, and the transition is monitored in transmission through the cavity.

The $CO_2$ frequency is measured by comparison with a very high-harmonic of the repetition rate of a femtosecond (fs) laser which, itself, generates a comb of frequencies [13, 14]. A second comb is produced by a sum-frequency generation (SFG) of the fs laser comb and the $CO_2$ laser in a nonlinear crystal. This results in a beat between the SFG comb and the high frequency part of the initial comb. The infrared frequency is thus compared to the difference between two modes of the comb. The beatnote is finally used to phase-lock the repetition rate to the $CO_2$ laser frequency. This scheme is independent of the comb offset and does not require any broadening of the comb.

The repetition rate is simultaneously compared to a 100 MHz or 1 GHz frequency reference, and the error signal is returned to the $CO_2$ laser via a servo loop of bandwidth of 10-100 mHz. The reference is generated at LNE-SYRTE and is based on a combination of a hydrogen maser and a cryogenic oscillator [15] controlled with a Cs fountain [16]. It is transferred to the LPL laboratory as an amplitude modulation on a 1.5 µm carrier, via 43 km of optical fibre [17]. The phase noise added by the fibre introduces an instability of a few $10^{-14}$ for a 1 s integration time, reducing to around $10^{-15}$ over 1000 s. These are figures for passive transfer, as normally employed here, but are improved more than 10 times when the fibre is compensated [17,18]. All radio frequency oscillators in the system are also referenced to the LNE-SYRTE signal.

The central Ramsey fringes are recorded by co-adding 10 frequency sweeps, alternating up and down, each of 20s. The spectrum, with a typical signal to noise ratio of 20-30 as illustrated in Figure 2, is then fitted to give the central frequency and the fringe periodicity. Thus, ultimately, the Cs and SF$_6$ frequencies are directly compared. The comparison reported here was carried out over a period of 2 years. During this period many parts of the experiment were refined, mainly in the frequency chain, and the SF$_6$ beam and the interferometer were completely dismantled and reassembled.

The mean value of the 487 individual measurements is:

$\nu$ (SF$_6$, P(4) E$^0$, central fringe) = 28 412 764 347 320.26±0.79 Hz, where the uncertainty (2.8×10$^{-14}$ as a fractional value) is the standard deviation. This value is lower than our earlier measurement, performed February-May 2003, by 2.5 Hz which is just 2$\sigma$ for that measurement [13]. The small deviation might be attributed to an excess of frequency noise in the interrogating laser which has been reduced before the current sets of measurements. The total range of the current measurement is less than ±2 Hz, that is 7×10$^{-14}$ as a fractional value, at which level a number of systematic effects might be relevant [19].

The light shift is +0.75±0.5 Hz and the slope is 0.06 Hz/mW at the $\pi/2$ pulse [13]. Thus, with a control of 10% in the laser power, this shift varies less than 0.1 Hz. A $\mu$-metal shield ensures that Zeeman shift variability is also below this level. The background pressure around the jet is typically 10$^{-7}$ mbar, which implies a negligible pressure shift [20].

The mean velocity and the velocity distribution of the molecular beam depend on both the pressure and temperature of the source. At the working pressure of 5 bar and a control level of 0.1 bar any pressure effect is less than 0.1 Hz. Temperature variation is more important, as its primary effect is to change the mean beam velocity. The second-order Doppler effect (SODE), which is -26 Hz, therefore changes. The fringe periodicity p also changes, because it is proportional to the velocity v : $p = \dfrac{v}{2D}$ where D is the distance between zones. Because of the unresolved hyperfine structure of the line, sets of fringes from different components overlap. This overlapping depends on the periodicity and affects the position of the central fringe. These systematic effects can be estimated from the correlation between the central frequency and the mean beam velocity or the periodicity. Frequency ν has been plotted as a function of both time t and periodicity p. A linear least-squares fitting has been done taking into account the experimental errors in periodicity and frequency, but with no error for the time variable. We obtain the dependence $\nu = a + b\,p + c\,t$ where the b coefficient gives the dependence of the central frequency on the periodicity, and the c coefficient gives the temporal variation of the frequency. With this analysis, we reduce the influence of the systematic effects discussed above. This is significant as it amounts to a factor approaching 2 in the temporal variation. The b-coefficient has a sign opposite to that of the SODE, but it results in a frequency variation of the same order of magnitude of the SODE variation. The frequency data can then be corrected for the periodicity dependence to leave just the dependence of frequency on time, as shown in Figure 3. Further, the blackbody radiation shift, which has never been estimated for a molecular transition around 10 μm, also depends on temperature. The temperature in the experimental room is not controlled better than ±3 K, and an estimation of the uncertainty due to all the temperature effects is 0.5 Hz.

The spectral characteristics of the local oscillator can affect the measurement. A degradation of the linewidth purity of the $CO_2$ laser induces a decrease in the fringe amplitude. Parasitic sidebands in the laser spectrum, for example due to mechanical vibrations or poor electromagnetic compatibility, affect both the position of the central fringe and the correct functioning of the control loops. The spectral width of the $CO_2$ laser was checked by recording its beatnote with another stabilised $CO_2$ laser. The effect of laser noise, as estimated from the beat and from the signature of the error signal, is less than 0.3 Hz on the central fringe.

As a last point, the uncertainty due to the simplified fitting model was estimated to less than 0.1 Hz. The quality of the primary data may be judged from Figure 2 which is quite typical.

The total uncertainty budget, including the frequency chain limitation is then 0.6 Hz or $2.2\times10^{-14}$ in fractional value.

The results of measurements over a two-year period are shown in Figure 3. The linear fit has a slope of $1.88\times10^{-14}$ per year (as a fractional value) with a statistical uncertainty of $0.12\times10^{-14}$. However, estimated uncontrolled systematic errors of $2.2\times10^{-14}$ in the measurements must be also taken into account (See also [13].). This induces a further error in the slope of $2.7\times10^{-14}\ yr^{-1}$. Thus, we place the upper limit on any variation of the relative $SF_6$ and Cs frequencies as $(1.9\pm0.12\pm2.7)\times10^{-14}\ yr^{-1}$.

From the equations above:

$$\frac{1}{(\nu(SF_6)/\nu(Cs))}\frac{\partial(\nu(SF_6)/\nu(Cs))}{\partial t}=-\frac{1}{2}\frac{1}{(m_P/m_e)}\frac{\partial(m_P/m_e)}{\partial t}-2.83\frac{1}{\alpha}\frac{\partial\alpha}{\partial t}-\frac{1}{(\mu_{Cs}/\mu_B)}\frac{\partial(\mu_{Cs}/\mu_B)}{\partial t}.$$

The interpretation of the measurement depends on which of the terms is considered to be constant or, in a model, their functional dependence. The result of [3] gives the frequency stability of atomic H compared with a Cs clock as $(3.2\pm6.3)\times10^{-15}\ yr^{-1}$, with exactly the same dependence on $\alpha,\mu_{Cs},\mu_B$ as here, which can thus be entirely removed. Data for the fractional temporal variation of $\alpha$ and $(\mu_{Cs}/\mu_B)$ can also be obtained from atomic clock experiments and, in both cases, the current limits are below $10^{-15}$ per year [2, 21]. Thus the current data implies a limit of $\frac{1}{(m_P/m_e)}\frac{\partial}{\partial t}(m_P/m_e)=(-3.8\pm5.6)\times10^{-14}\ yr^{-1}$, a conclusion which is independent of any model. Another model-independent figure is $(-0.4\pm1.3)\times10^{-11}\,yr^{-1}$ obtained indirectly from a measurement of the Rydberg [7].

Limits may also be placed on $\frac{1}{(m_P/m_e)}\frac{\partial}{\partial t}(m_P/m_e)$ from astronomical observations. By comparing wavelengths in electronic spectra of $H_2$ as measured locally on the earth and the corresponding wavelengths from selected quasars a significant result was claimed: $(-1.7\pm0.5)\times10^{-15}\,yr^{-1}$ [10]. More recently, however, comparison of ammonia inversion lines and other molecules in quasars yielded a lower limit with a null result: $(-1\pm3)\times10^{-16}\,yr^{-1}$ [11].

Limits on $\frac{1}{(m_P/m_e)}\frac{\partial}{\partial t}(m_P/m_e)$ can alternatively be inferred from laboratory comparisons of atomic clocks and the most stringent figure from this type of comparison is again null: $(-1.2\pm2.2)\times10^{-15}\,yr^{-1}$ [22]. However, this is not a direct result as the Schmidt model must be invoked for the nuclear magnetic moment [7].

There are thus now three routes to the proton-electron mass ratio, probing different areas of Physics. From these, only the results reported here are direct and model-free. All the laboratory comparisons enjoy the advantage of controlled environments so that high precision becomes available in each measurement, while systematic effects can be extensively studied and reduced. The time scales are, however, short. By contrast, astrophysical data are obtained from environments over which there is no control. There is lower frequency accuracy but measurements are effectively separated in time by several Gyr. Given the difficulties in the theoretical and experimental backgrounds it is particularly important to explore the very different time scales, using complementary techniques to give reliability to the results.

The authors are grateful to the LNE-SYRTE for providing the reference signal from their primary standard for the absolute frequency measurements by means of the optical link between the two laboratories. A. Goncharov, O. Lopez and C. Daussy have given unstinting support to this project over many years, and we are very grateful for their efforts.

* E-mail : amy@univ-paris13.fr

† Permanent address: P.N.Lebedev Physical Institute, Leninsky Prospect, 53, Moscow, 119991, Russia.

‡ Permanent address: The Cavendish Laboratory, Madingley Road, Cambridge CB3 OHE, UK.

**Figure captions**

FIG. 1: Experimental set up.

FIG. 2: Fringes at 200 Hz, obtained using a 1m interzone separation. Experimental conditions: pure $SF_6$ beam, input pressure $5 \times 10^5$ Pa, 12 mW inside U cavity FM modulation at 115 Hz index 0.43, 75 µW inside the detection cavity, time constant for detection 0.1s. Average of 5 up-down sweeps, 200 points, averaging 1s per point. Signal-to-noise ratio 30.

FIG. 3: Absolute frequency of the central fringe displayed as a function of time. The y-axis is offset by 28 412 764 347 000 Hz. The least squares best fit line has a slope of $1.88 \times 10^{-14}$ /yr.

Figure 1

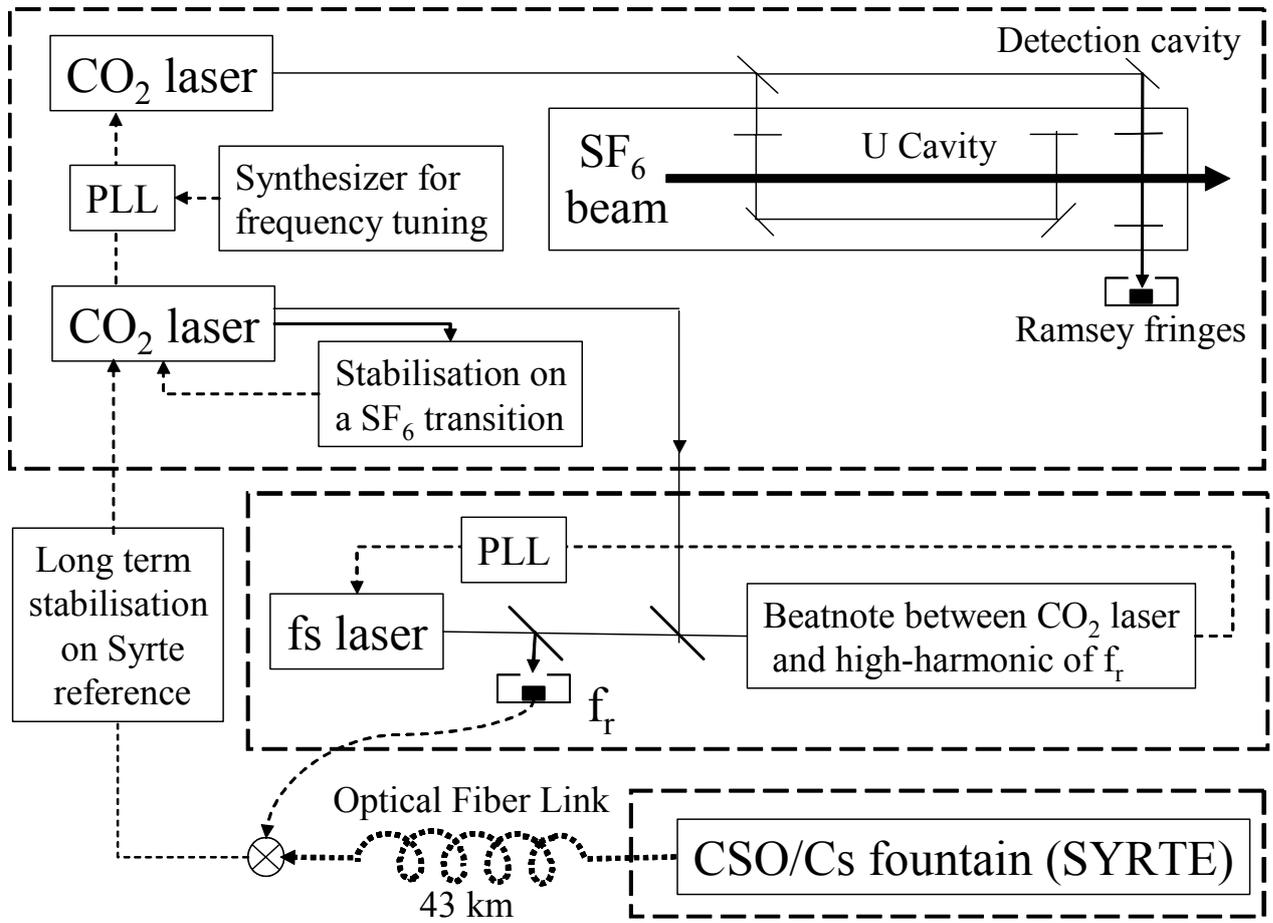

Figure 2

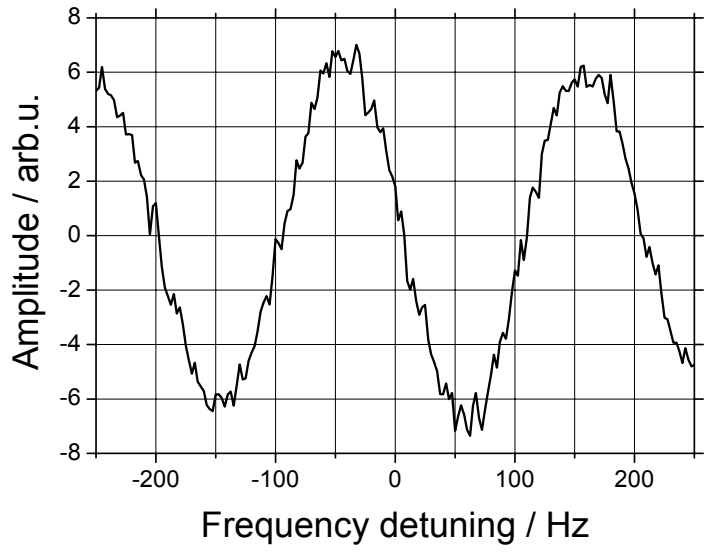

Figure 3

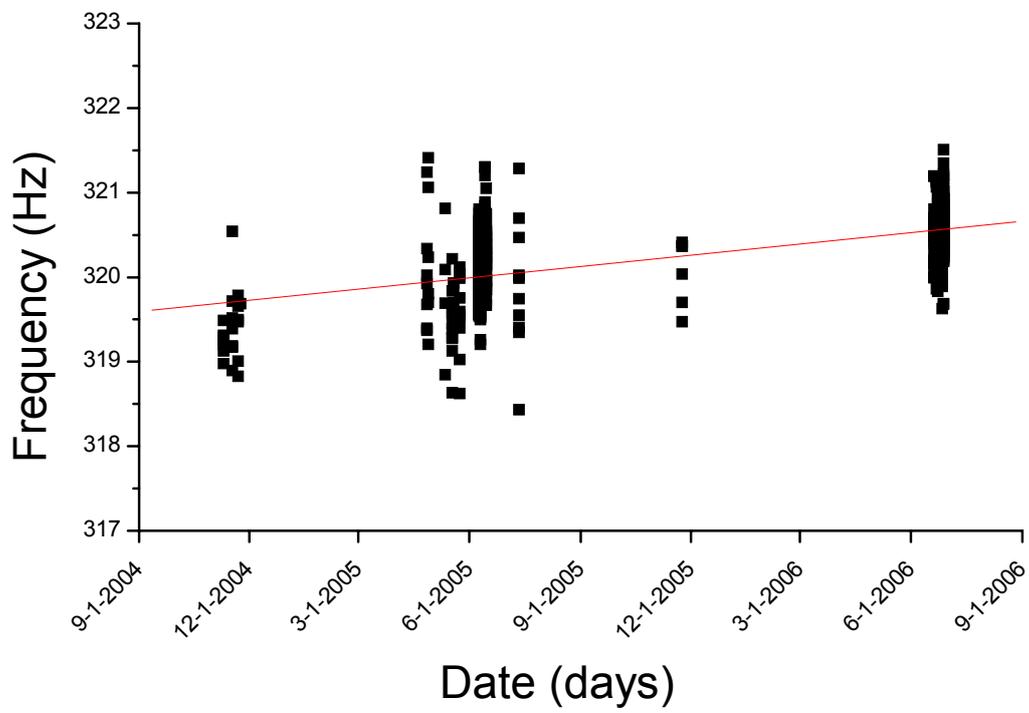